\documentclass[12pt] {article}

\usepackage{amssymb}
\usepackage{epsfig}

\def\gev{$\mbox{GeV}^2$}

\begin{document}

\setlength{\baselineskip}{0.75cm}
\setlength{\parskip}{0.45cm}
\begin{titlepage}
\begin{flushright}
\large
DO-TH 96/18\\ September 1996
\end{flushright}
\normalsize
\vspace{1.3cm}
\begin{center}
\LARGE
{\bf Small $x$ resummations confronted with $F_2(x,Q^2)$ data}\\
\vspace{1.3cm}
\large
I.\ Bojak and M.\ Ernst\\
\vspace{1cm}
Institut f\"ur Physik, Universit\"at Dortmund \\
D--44221 Dortmund, Germany
\end{center}
\vspace{1.8cm}
\underline{{\large{Abstract}}} \\[2ex]
\normalsize
\setlength{\baselineskip}{0.75cm}
\setlength{\parskip}{0.45cm}
It has been observed recently that a consistent LO BFKL gluon evolution
leads to a steep growth of $F_2(x,Q^2)$ for $x\rightarrow 0$ almost
independently of $Q^2$.
We show that current data from the DESY HERA collider are precise enough
to finally rule out a pure BFKL behaviour in the
accessible small $x$ region.
Several attempts have been made by other groups to treat the BFKL
type small $x$ resummations instead as additions to the conventional anomalous
dimensions of the successful renormalization group ``Altarelli-Parisi''
equations. We demonstrate that {\em all} presently available
$F_2$ data, in particular at lower values of $Q^2$, can {\em not}
be described using the presently known NLO (two-loop consistent)
small $x$ resummations.
Finally we comment on the common reason for the failure of these BFKL
inspired methods which result, in general, in too {\em steep}
$x$-dependencies as $x\rightarrow 0$.  

\begin{center}
(To be published in Phys.\ Lett.\ B)
\end{center}
\end{titlepage}
\newpage
%
%
%
In spite of the success \cite{HERAold, HERA93, HERA94} 
of the conventional renormalization
group ``Altarelli-Parisi'' equations (RGE) \cite{ap} much
work has focused recently on replacing or modifying them by small $x$
resummations. It should be stressed that, so far, this is not required by
the experimental data. The rise of $F_2$ at small $x$, for
example, has been predicted by the GRV group \cite{grv90,grv} using the
NLO RGE and thus is not a unique sign of the BFKL \cite{bfkl} type of growth.
By using the BFKL evolution methods of Askew et al.\ \cite{AKMS}
consistently, we did even show in a previous paper \cite{oldpap}
that it is probable that a pure LO BFKL evolution is {\em not}
compatible with the observed growth in the $F_2(x,Q^2)$ data. We confirm
this here using more recent data from the DESY HERA collider
\cite{HERA93,HERA94}.

But this does not necessarily prove that the BFKL resummation of
leading logarithms in $x$ is inadequate. The original resummation
ignores the quarks and uses a fixed strong coupling constant.
Thus it is expected to be relevant only in the {\em asymptotic}
small $x$ limit and it could still be valid at very small values
of $x$ that are currently not accessible by experiment.
There is also the possibility that the corresponding NLO resummations
will change the LO behaviour and that the assumption of gluon
dominance limits the region of applicability strongly. 

Keeping this in mind the ``high energy'' ($s \gg Q^2$ leading to small $x$
via $x\sim Q^2/s$) resummations done by Catani et al.\ \cite{catani}
seem to offer a way out. They treat the BFKL type resummations
as additions to the fixed order anomalous dimensions and Wilson
coefficients, hoping to obtain an improved perturbative expansion
that way. They also resummed the NLO small $x$ contributions in the quark
sector.
Furthermore, they calculated all resummations consistently by using the
$k_T$-factorization and showed that their methods do not
spoil the all-order factorization of collinear singularities.
We note that by recalculating the LO resummations they have confirmed
the known BFKL results.

The obvious question is then of course if the ``improved'' expansion
can match the success of the conventional RGE. First comparisons
with $F_2$ data using these methods were made by Ellis et al.\
\cite{ehw}. They stressed the importance of enforcing the
fundamental energy-momentum conservation which is generally broken
by the resummations. Starting the evolution of flat input densities
from $Q_0^2=4$ \gev ,
they can describe the $F_2$ data fairly well. In this paper
we will show that the agreement is spoiled when the starting
scale is lowered to values typical for recent RGE fits,
in order to take into account {\em all} presently available
HERA data.  
 
First we will present an update of the analysis of $F_2$ data
using the LO BFKL evolution of the gluon. We give now a brief
outline of the methods employed in \cite{oldpap}, noting that
the references for the formulae and methods used can also be
found there. The BFKL equation concerns the unintegrated gluon
density $f(x,k^2)$, because the transverse momentum $k^2$
of the gluons in the corresponding ladder diagram is not strongly ordered
as in the LO RGE case and cannot be integrated out.

The connection between the integrated and unintegrated gluon
is given by the equations
\newcounter{temp}
\setcounter{temp}{\value{equation}}
\stepcounter{temp}
\def\theequation{\arabic{temp}\alph{equation}}
\setcounter{equation}{0}
\begin{eqnarray}
\label{fg_int}
x\, g(x,Q^2)&=&\int_0^{Q^2}\,\frac{dk^2}{k^2}\, f(x,k^2),\\
\label{fg_diff}
f(x,k^2)&=&\left.\frac{\partial x\, g(x,Q^2)}{\partial\ln Q^2}
\right|_{Q^2=k^2},
\end{eqnarray}
\setcounter{equation}{\value{temp}}
\def\theequation{\arabic{equation}}
and in terms of the unintegrated gluon one can then write the BFKL
equations as an evolution equation in $x$
\begin{equation}
\label{master}
-x\,\frac{\partial\, f(x,k^2)}{\partial x}=\frac{3\,\alpha_s (k^2)}{\pi}
\, k^2\int\,\frac{dk'^2}{k'^2}\,\left[\frac{f(x,k'^2)-f(x,k^2)}{|k'^2-k^2|}+
\frac{f(x,k^2)}{(4\, k'^4+k^4)^\frac{1}{2}}\right].
\end{equation}

Predictions for $F_2$ are then made by convoluting the BFKL
gluon with the photon-gluon fusion quark box $F^{(0)}$ using the
$k_T$-factorization theorem
\begin{equation}
\label{kt}
F_i(x,Q^2)=\int\,\frac{dk'^2}{k'^4}\,\int_x^1\,\frac{dy}{y}\,
f\left(\frac{x}{y},k'^2\right)\, F_i^{(0)}(y,k'^2,Q^2),
\end{equation}
with $i=T,L$ denoting the transverse and longitudinal parts, respectively.
Since we have completely ignored the quark sector and non-perturbative effects,
we will add a background fitted to the data. Obviously the background
should not dominate over the BFKL contribution and it also should not have
a strong $x$ dependence that would mask the BFKL behaviour. We use
a soft Pomeron ansatz $C_{I\! P}\, x^{-0.08}$, where the constant $C_{I\! P}$
is fitted separately for each $Q^2$ bin. It is well known that
the BFKL equation has the same double logarithmic limit as the LO RGE,
when its (fixed) coupling constant $\alpha_s$ is replaced by the standard
running one of the LO RGE. Since we want to connect to the LO RGE
at large $x$, this replacement has to be made in (\ref{master}).

LO RGE gluon distributions are then used at
$x_0=10^{-2}$ as input via Eq.~(\ref{fg_diff}). We now have to
face the difficulty that (\ref{master}) requires knowing
$f(x,k^2)$ down to $k^2=0$ \gev, but of course even the special
LO RGE gluons $g^{\mathrm{RGE}}(x,Q^2)$ used in \cite{oldpap}
do not extend far enough down
in $Q^2$ to supply that information. Thus we are forced to make
an ansatz for the infrared (IR) region, if we want to avoid cutoffs
to the integral. The introduced running coupling also needs
to be taken care of in the region of small $k^2$. Following Askew et al.\
\cite{AKMS}, the IR ansatz of \cite{oldpap} is
\begin{equation}
\label{iransatz}
f(x,k^2)=\frac{k_c^2+k_a^2}{k_c^2}\frac{k^2}{k^2+k_a^2} f(x,k_c^2)\quad
\mathrm{for}\quad k^2<k_c^2.
\end{equation}
At $x=x_0$ we additionally shift the gluon
\begin{equation}
\label{irshift}
	f^{\mathrm{RGE}}(x_0,k^2) \longrightarrow f^{\mathrm{RGE}}
	(x_0,k^2+k_a^2) \label{eqshift}.
\end{equation}
The strong coupling is frozen: $\alpha_s(k^2) \longrightarrow
 \alpha_s(k^2+k_b^2).$ The ansatz is scrutinized closely in
\cite{oldpap}. We just want to remind the reader of the salient
points: old gluon distributions have to be used, since most of the
more recent ones already create a steep growth of $F_2$ in the low
$x$ region without any additional BFKL boost. The dependence
of $F_2$ on the IR parameters is under control, except for
$k_a^2$ which has to be fixed by forcing the
integrated gluon of (\ref{irshift}) to match the original
gluon $g^{\mathrm{RGE}}$ at medium to high $Q^2$.
For the $D_0$-type gluon used below we obtain $k_a^2=0.95$ \gev\
\cite{oldpap}. The strong BFKL growth of $F_2$ could be suppressed \cite{AKMS} by increasing
 $k_a^2$, but this would of course spoil the consistency with the RGE gluon.

Parametrizations of our results were given in \cite{oldpap}.
Using the $D_0$-type gluon parametrization and fitting the background
to the new data below $x=10^{-2}$ from
the HERA collaborations \cite{HERA93,HERA94} at DESY as well as to
data from the Fermilab E665 experiment \cite{E665}, we obtain
the solid curve in Fig.~\ref{fig1}. The background contribution
is shown as dashed curve. Note that at $Q^2=3.5$ \gev\ and at $Q^2=6.5$
\gev\ $C_{I\! P}$ would be {\em negative} and is set to zero by hand.
It is obvious
that in spite of having the freedom of fitting $C_{I\! P}$, the data
can {\em not} be described by the BFKL curves. The steep growth of 
$F_2\sim x^{-0.5}$ predicted by the BFKL equation is simply too strong,
especially at medium to low $Q^2$. Thanks to the precision of the
new $F_2$ data pure BFKL evolution is now definitely ruled out!

Of course immediately the question arises if this can be ameloriated
by using modified evolution equations that incorporate the
successful RGE. That the NLO RGE are very
successful indeed at describing the $F_2$ data is made obvious by
the dotted line in Fig.~\ref{fig1} which shows the R1 fit of MRS
\cite{MRSR}. We are especially interested in elucidating the question
if modified equations can match the success of the NLO RGE in the
region of low to medium $Q^2$. The fact that even below $Q^2=4$ \gev\
data can be described using those equations has been pioneered
by the GRV group \cite{grv90,grv} and is now becoming universally accepted
\cite{HERA93,HERA94,MRSR,CTEQ4}. This region is also expected to lead
to difficulties with the resummed BFKL pole at 
$\lambda=\frac{3\,\alpha_s}{\pi}\,4\,\ln 2$, because
$\alpha_s$ increases strongly in this region if it runs.

We have tried the method employed by Forshaw et al.\ \cite{forshaw}.
They limit their calculations to the LO small $x$ resummations of the
gluon by replacing the standard anomalous dimension in the RGE by the
resummed one and by ignoring the quarks. The (unsuccessful)
results of trying to extend this method down to lower $Q^2$ will be
reported in a forthcoming detailed article \cite{review}.
Instead we concentrate
here on the method developed by Ellis et al.\ \cite{ehw}. 
Following the work of Catani et al.\ \cite{catani} they add the known
LO and NLO resummation corrections to the NLO RGE singlet anomalous
dimensions and to the Wilson coefficients.
Note that the corrections to the non-singlet anomalous
dimensions are less singular than $1/x$ for $x \rightarrow 0$ and
need not be taken into account. We use
the Mellin transformation with respect to $x$ in order to express the results
in terms of moments
\begin{equation}
\label{mellin}
f(n,Q^2)=\int_0^1\, dx\, x^n\, f(x,Q^2),
\end{equation}
where the Mellin moment is shifted by one compared to the usual
convention $(x^{n-1})$, since this will allow a more compact notation in the
following.

To the fixed order expansion of the anomalous dimensions
\begin{equation}
\label{fix}
\gamma (n,\alpha_s)=\frac{\alpha_s}{2\,\pi}\, \gamma^{(0)}(n)+
\left(\frac{\alpha_s}{2\,\pi}\right)^2\, \gamma^{(1)}(n)+
{\cal O}(\alpha_s^3)
\end{equation}
towers of small-$x$ resummation corrections are now added.
In LO, that is summing terms of the order $(\alpha_s/n)^k$, only corrections
to the anomalous dimensions $\gamma_{g\Sigma}$ and $\gamma_{gg}$ exist,
where $\Sigma\equiv\sum_q\, (q+\overline{q})$ denotes the singlet
quark density.
In NLO the  corrections to $\gamma_{\Sigma\Sigma}$ and
to $\gamma_{\Sigma g}$ are known, but those of the gluon
have not yet been completed.
The additional resummation corrections can thus compactly be written as
\begin{eqnarray}
\label{corrections}
\hat{\gamma}_\mathrm{res.}&=&
\left(\begin{array}{cc} \gamma_{\Sigma\Sigma}&\gamma_{\Sigma g}\\
\gamma_{g\Sigma}&\gamma_{gg}\end{array}\right)_{res.}\nonumber\\
&=&\left(
\begin{array}{cc} 0&0\\ \frac{C_F}{C_A}\,\gamma_{L}&\gamma_{L}\end{array}
\right)+\left(
\begin{array}{cc} 2\, n_f\,\frac{C_F}{C_A}\,\left[\gamma_{NL}-\frac{2\,
\alpha_s}{3\,\pi}\,T_R\,N_f\right]&2\, n_f\, \gamma_{NL}\\0\, (?)&0\,
(?)\end{array}
\right)+\\
&&+{\cal O}\left(\alpha_s^2\,\left(\frac{\alpha_s}{n}\right)^k\right),
\nonumber
\end{eqnarray}
with $k>0$, and the question marks signify the yet unknown entries.

To obtain the BFKL gluon anomalous dimension $\gamma_L$ from
\begin{equation}
n=\overline{\alpha}_s \chi (\gamma_L);\quad
\overline{\alpha}_s\equiv\frac{C_A\,\alpha_s}{\pi};\quad
\chi (\gamma)\equiv 2\,\psi (1)-\psi (\gamma )-\psi(1-\gamma ),
\end{equation}
we use a complex Newton method. For the NLO resummation
obviously the choice of scheme is important. Since up to now
completely resummed expressions are only available in the DIS factorization
scheme, we use this scheme in the following.

The DIS scheme is defined by $C_2^\Sigma=1$ and $C_2^g=0$, thus all the
resummation corrections
are shifted into the anomalous dimensions. Then we can write
\begin{eqnarray}
\gamma_{NL}&=&\frac{1}{2\, n_f}\, h_2(\gamma_L ,n)\, R(\gamma_L),\\
h_2(\gamma,n)&=&h_2(\gamma )\, (1+{\cal O}(n))\quad\mathrm{for}
\quad n\rightarrow 0,\\
h_2(\gamma )&=&2\, n_f\,\frac{\alpha_s}{2\,\pi}\, T_R\,
\frac{2+3\,\gamma-3\,\gamma^2}{3-2\,\gamma}\,
\left(\frac{\pi^2\, \gamma^2}{1-4\,\gamma^2}\,
\frac{1}{\sin (\pi\,\gamma)\,\tan (\pi\,\gamma)}\right),\\
R(\gamma )&=&\left\{\frac{\Gamma (1-\gamma )\,\chi (\gamma )}
{\Gamma (1+\gamma )\, \left[-\gamma\,\chi '(\gamma )\right]}\right\}
^\frac{1}{2}\,\exp\left\{\gamma\,\psi (1)+\int_0^\gamma\, d\gamma\,
\frac{\psi '(1)-\psi '(1-\gamma)}{\chi (\gamma )}\right\}.
\end{eqnarray}
Here the prime denotes the first derivative with respect to $\gamma$.
The expressions can be directly
calculated once $\gamma_L$ has been determined.

It is a non-trivial check of the work of Catani et al.\ that
by expanding the small $x$ resummations in $\overline{\alpha}_s/n$
one obtains the same constant respectively $1/n$ singular terms
as in the (NLO) RGE in the $n\rightarrow 0$ limit. To avoid double-counting,
these terms have to be subtracted before the additional resummations are added.
Effectively, Eq.~(\ref{corrections}) is then replaced by
\begin{eqnarray}
\label{effcorr}
\hat{\gamma}_\mathrm{res.}&=&
\left(
\begin{array}{cc} 2\, n_f\,\frac{C_F}{C_A}\,\gamma_q&2\, n_f\,\gamma_q\\
\frac{C_F}{C_A}\,\gamma_g&\gamma_g\end{array}
\right)\\
\gamma_g&=&\gamma_L-\frac{\overline{\alpha}_s}{n}\\
\gamma_q&=&\left[h_2(\gamma_L)\,R(\gamma_L)-
\frac{\alpha_s}{2\,\pi}\, T_R\,
\frac{2}{3}\,\left(1+\frac{13}{6}\,\frac{\overline{\alpha}_s}{n}\right)
\right].
\end{eqnarray}
Here we can see clearly that the corrections to the splitting into
gluons and quarks just differ by a colour factor.
It should also be pointed out, that the first terms in the expansions
of the effective corrections are
\begin{eqnarray}
\label{effexp}
\gamma_g&=&2\,\zeta (3)\,\left(\frac{\overline{\alpha_s}}{n}\right)^4+
{\cal O}\left(\left(\frac{\overline{\alpha}_s}{n}\right)^6\right),\\
\label{effexp2}
\gamma_q&=&\frac{\alpha_s}{2\,\pi}\, T_R\,\frac{2}{3}\,\left[
\left(\frac{71}{18}-\zeta (2)\right)\,\left(\frac{\overline{\alpha}_s}{n}
\right)^2+{\cal O}\left(\left(\frac{\overline{\alpha}_s}{n}\right)^3
\right)\right]
\end{eqnarray}
so that the correction in the quark sector leads in terms
of powers of $\alpha_s$ and also starts with the lower power of $1/n$.
Thus we expect the quark correction to play an important role
despite its being formally subleading. The quark correction is also
expected to vanish more slowly for $x\rightarrow 1$, because it
does not fall off as steeply when $n\rightarrow\infty$.

It is of course entirely possible that the NLO order corrections in
the gluon sector change those expectations, since they depend on the fact
that the second and third coefficients in the expansion of $\gamma_L$
vanish. 
A further problem is that the added corrections violate the fundamental
energy-momentum conservation, as can be easily checked by setting
$n\rightarrow 1$. The calculation of higher resummation corrections
will not solve this problem, since the LO corrections cannot be
cancelled. Of course the original RGE manifestly obey energy-momentum
conversation. This is not only a theoretical problem, since the
unconserved growth is very strong and leads to obviously false
results. We will look at the simplest ways to restore the conservation
of energy and momentum. The ``hard'' way \cite{ehw} is simply to
subtract the troublesome values of the corrections via
$\hat{\gamma}_{res.}(n)\rightarrow\hat{\gamma}_{res.}(n)-
\hat{\gamma}_{res.}(1)$. As has already been remarked in \cite{ehw}
this does not much reduce the growth and it has turned out to be
impossible to fit the experimental $F_2$ data using this approach.
Thus we will limit ourselves to the second ``soft'' method,
multiplying the corrections by a factor that goes to zero for
$n\rightarrow 1$.

This conserving factor, being introduced in an entirely ad hoc way,
should go to unity in the limit of very small $n$ which
corresponds to small $x$. We test the following factors: $(1-n)$,
$(1-n)^2$ and $(1-2\, n+n^3)$. They have been studied previously
in \cite{blume} using higher starting scales $Q_0^2$.
Obviously these factors will inhibit
the growth more strongly, since they weaken the singularities as
can be seen by looking at the expansions (\ref{effexp}) and (\ref{effexp2}).
It should also be noted that multiplying the quark correction by the third
factor leads to a term $\sim n$. This means that the correction
times this factor cannot be transformed back into $x$-space on its own.
Instead it needs a parton distribution that falls off at least with
$1/n$. While this is fulfilled for all reasonable parton distributions,
it is still a hint that the parton distributions are not correctly
factorized anymore. For this reason we do not examine even higher
powers of $n$ in the conserving factor. By checking this range of
factors we also get an estimate of the influence of the missing higher order
corrections, since they will also be suppressed in powers of $1/n$ versus
powers of $\overline{\alpha}_s$.

The RGE are only known to two loops and can be conveniently solved
in $n$-space. We wish to modify the known NLO RGE $n$-space
solution \cite{grv90,furm},
but then it is necessary to check if the corrections can be
added to the two-loop solution consistently. Since we only know the
running coupling to two loops, we can write the evolution equation
as
\begin{equation}
\frac{d\vec{q}(n,Q^2)}{d\alpha_s}=\left[\frac{\alpha_s}{2\,\pi}\,
\hat{\gamma}^{(0)}(n)+\left(\frac{\alpha_s}{2\,\pi}\right)^2\,
\hat{\gamma}^{(1)}(n)+\hat{\gamma}_{res.}\right]\,
\frac{1}{-\frac{\beta_0}{4\,\pi}\,\alpha_s^2-\frac{\beta_1}{(4\,\pi )^2}\,
\alpha_s^3}\, \vec{q}(n,Q^2),
\end{equation}
using for the strong coupling $(n_f$ is the number of active
massless quark flavours$)$
\begin{equation}
\frac{\alpha_s(Q^2)}{4\,\pi}=\frac{1}{\beta_0\,\ln (Q^2/\Lambda^2)}-
\frac{\beta_1\,\ln\ln(Q^2/\Lambda^2)}{\beta_0^3\, [\ln(Q^2/\Lambda^2)]^2},
\quad\beta_0=11-\frac{2}{3}\, n_f,\quad\beta_1=102-\frac{38}{3}\, n_f.
\end{equation}
In \cite{review} it is shown that a consistent addition of the corrections
can be accomplished by replacing
\begin{equation}
\label{conadd}
\hat{\gamma}^{(1)}(n) \longrightarrow \hat{\gamma}^{(1)}(n) +
\frac{\pi\,\beta_0}{\alpha_s(Q^2)-\alpha_s(Q_0^2)}\,\int_{\alpha_s(Q_0^2)}
^{\alpha_s(Q^2)}\, d\alpha\,\frac{\hat{\gamma}_{res.}(n,\alpha )}
{\frac{\beta_0}{4\,\pi}\,\alpha^2+\frac{\beta_1}{(4\,\pi )^2}\,\alpha^3}.
\end{equation}
We have tested that by using this method we can reproduce the results
shown in \cite{ehw}.

The dashed curve of Fig. \ref{fig2} shows the strong effects of the added
resummations compared to the solid curve of the conventional RGE. 
For both curves the unchanged R1 input distributions at $Q_0^2$ were
used and the resummations were added using the conserving factor $(1-n)$. 
Obviously it is necessary to refit the input parton distributions,
since the growth in the small $x$ region is incompatible with the data.
It is interesting to note the similarity of that growth to the
LO BFKL one shown in Fig.~\ref{fig2} as dotted line. This is a first
hint that fitting experimental data with the new method
will lead to the same problems at small $x$. 
In order to test if the RGE plus corrections can fit the experimental
data as well as the pure RGE, we would in principle need to fit to
the same range of experiments as is done for example in \cite{MRSR}.
Even if the resummation corrections to all those processes were known,
the computing time would still be prohibitive due to the complicated
calculation of the corrections.

Instead we use the following approach to save computing time:
we take the optimal MRS R1 \cite{MRSR} fit as the basis for our fit by
using the same
starting scale $Q_0^2$, $\Lambda_{QCD}(n_f=4)$ and the same valence
distributions for the quarks. Since the R1 fit describes the HERA data
very well, we use it with the unmodified RGE to generate $F_2$ points
for our fit. Below $x=5\cdot 10^{-2}$ the errors and $x$-range used
for the points correspond to those of the HERA data. Above this value of $x$
we force our fit to reproduce the R1 fit by setting an artificial
error of one percent. In this large $x$ region not only $F_2$ but also
the sea and the gluon are tested. This simulates the numerous experiments
constraining the parton distributions at large $x$ which are taken into
account by the R1 fit.
Since we do our calculations in the DIS scheme, we have to apply the
corrections for the change of scheme to the $\overline{\mathrm{MS}}$
partons of R1. This is done at the starting scale $Q_0^2=1$ \gev\ and
then the RGE in the DIS scheme are used to obtain the constraints.
In fact the fitting program uses the same transformation, so that
the new fitted parameters are valid in the $\overline{\mathrm{MS}}$ scheme
and can be directly compared to those of the MRS R1 fit.

It turns out that the modified evolution equations cannot reproduce well the
RGE sea in the large $x$ region. The gluon and $F_2$ are reproduced
with reasonable accuracy. But the sea is less and less in agreement
when the maximum power of $n$ in the conserving factor is raised.
This can be easily understood by noting that the product of the
conserving factor with the expansion (\ref{effexp2}) does not fall off
as quickly for $n\rightarrow\infty$ as the product with (\ref{effexp})
does. Relaxing the strong constraints
at large $x$ is not a viable solution to the problem. This is so because,
as we will see, the $F_2$ data at small $x$ are not well described by
the modified evolution equations. Thus once the tight constraints
in the large $x$ regions are removed, the fit program adjusts the parton
distributions for the small $x$ region and basically ignores any
weak constraints imposed at large $x$. This leads to strong deviations
especially of the sea from the original R1 partons. Since here our
goal is not to give an alternative description of the small $x$ data,
but rather to check if modified evolution equations can match the RGE
in describing {\em all} data, we cannot avoid imposing the constraints
at large $x$.

The results of our fits are displayed in
Fig.~\ref{fig3}. The curve denoted as ``no conservation'' is the fit without
conversation of energy and momentum, corresponding to the original
ideas of Catani et al.\ \cite{catani}. The one described as
``shift'' is a fit using the conserving factor $(1-n)$ but also
shifting the strong coupling using $\alpha_s(Q^2)\rightarrow
\alpha_s(Q^2+1.5 \,\mathrm{GeV}^2)$. We note that the fit using
the conserving factor $(1-2\,n+n^3)$ is not displayed. It leads to
an unacceptable fit which oscillates around the low
$x$ data strongly, especially at the low $Q^2$ values. This confirms
our earlier suspicions about possible problems with this factor.
The ``no conservation'' curve shows that the unconserved growth is way too
strong to be in accord with experimental data. 

It is also easy to see that increasing the maximum power of
$n$ in the conserving factor leads to the expected  decrease of the growth
in the small $x$ region. Still it is obvious that the curve
corresponding to the $(1-n)$ factor cannot describe the data
except perhaps at the highest $Q^2$ bin. The curve pertaining to
the $(1-n)^2$ factor offers some improvement. This is expected
due to the better cancellation of the leading term in the
added quark resummation (\ref{effexp2}) for $n<1$ which leads
to a smaller deviation from the standard RGE. But the small $x$ region
below $12$ \gev\ is still not fitted well. That this is caused by
the strong BFKL like growth at small $x$ is especially
obvious at $1.5$ \gev . Here we can see that the fit has
adjusted the parton distributions so that the curves at medium to
low $x$ are below the data to compensate the growth at very
small $x$ somewhat.

Looking at the ``dips'' in the curves we can also
clearly see that the onset of the growth is shifted towards lower
$x$ when the power of $n$ in the conserving factor is increased.
We can conclude from these fits that the effects of changing
the conserving factor are sizable in the small $x$ region and that
no definite predictions on $F_2$ at small $x$ can be made for
the time being using modified evolution equations. On the other
hand the observed strong rise at small $x$ could be taken as an indication
that the low
$Q^2$ region will not be better described once all NLO terms of the
resummations compatible with the two-loop RGE are known\footnote{
However higher order terms introduced by a purely kinematic constraint on the
gluon ladder suggest that the growth of the gluon
could be reduced \cite{KMS}.}.
The BFKL like growth simply
is {\em too strong} for low $Q^2$ and spoils agreement with data
(rather flat in $x$) whenever
it becomes dominant. This is shown nicely by the curve labeled ``shift''
which is generated using a shifted value of $\alpha_s$. Obviously
the fit to the data is much improved at low $Q^2$ since the
BFKL growth is suppressed due to the artificially lowered value
of $\alpha_s$.

All our attempts to describe experimental data of $F_2$ using methods
inspired by the BFKL equations have failed. The direct application
of the BFKL equation in the small $x$ region is now finally ruled
out by experimental data \cite{HERA93,HERA94,E665}. Adding the small $x$
resummations to the RGE does not lead to a conflict with data, if the parton
distributions are adjusted accordingly and provided the starting scale
$Q_0^2$ is kept around $4$ \gev\ \cite{ehw}, i.\ e.\ if only a fraction
of all available data is taken into account.
But it is by now widely accepted that it
is possible to describe data below this $Q_0^2$ scale using the NLO
RGE. The modified evolution equations fail this acid test. This conclusion can
be drawn despite the apparent difficulty to obtain any definite predictions
using these equations at all.

A further indication of the reliability of our results is provided by a similar
analysis of Ball and Forte \cite{bafo}. They also found that deviations
from the standard RGE are not supported by recent HERA data, although
their method differs in some respects from ours. Notably they switch on
the small $x$ resummations at a fitted $x_0$, they estimate the missing
NLO gluon small $x$ contributions by using energy-momentum conservation,
and refit {\em only} the small $x$ tails of given parton distributions.  

There is a common reason for these failures. At low $Q^2$ the BFKL
pole $\lambda=\frac{3\,\alpha_s}{\pi}\,4\,\ln 2$ dominates over all the
other poles since it moves strongly along the positive real axis in the
$n$ plane due to the growth of $\alpha_s$. The validity of this argument
is nicely demonstrated by the curve labeled ``shift'' in Fig. \ref{fig3}.
Simply calculating the values for $\lambda$ at $Q^2=4$ \gev\
and at $Q^2=1$ \gev\ suggests that any future method that includes
this pole in the evolution is likely to have trouble to describe the moderate
rise in $x$ of $F_2(x,Q^2)$ observed in the experimental data
at lower values of $Q^2$.

{\bf Acknowledgments:} We thank E.\ Reya and M.\ Gl\"uck for suggestions
and instructive discussions. This work has been supported in part by the
'Bundesministerium f\"ur Bildung, Wissenschaft, Forschung und Technologie',
Bonn.

\newpage
%
%
%
\section*{Figure Captions}
\begin{figure}
\begin{center}
\epsfig{file=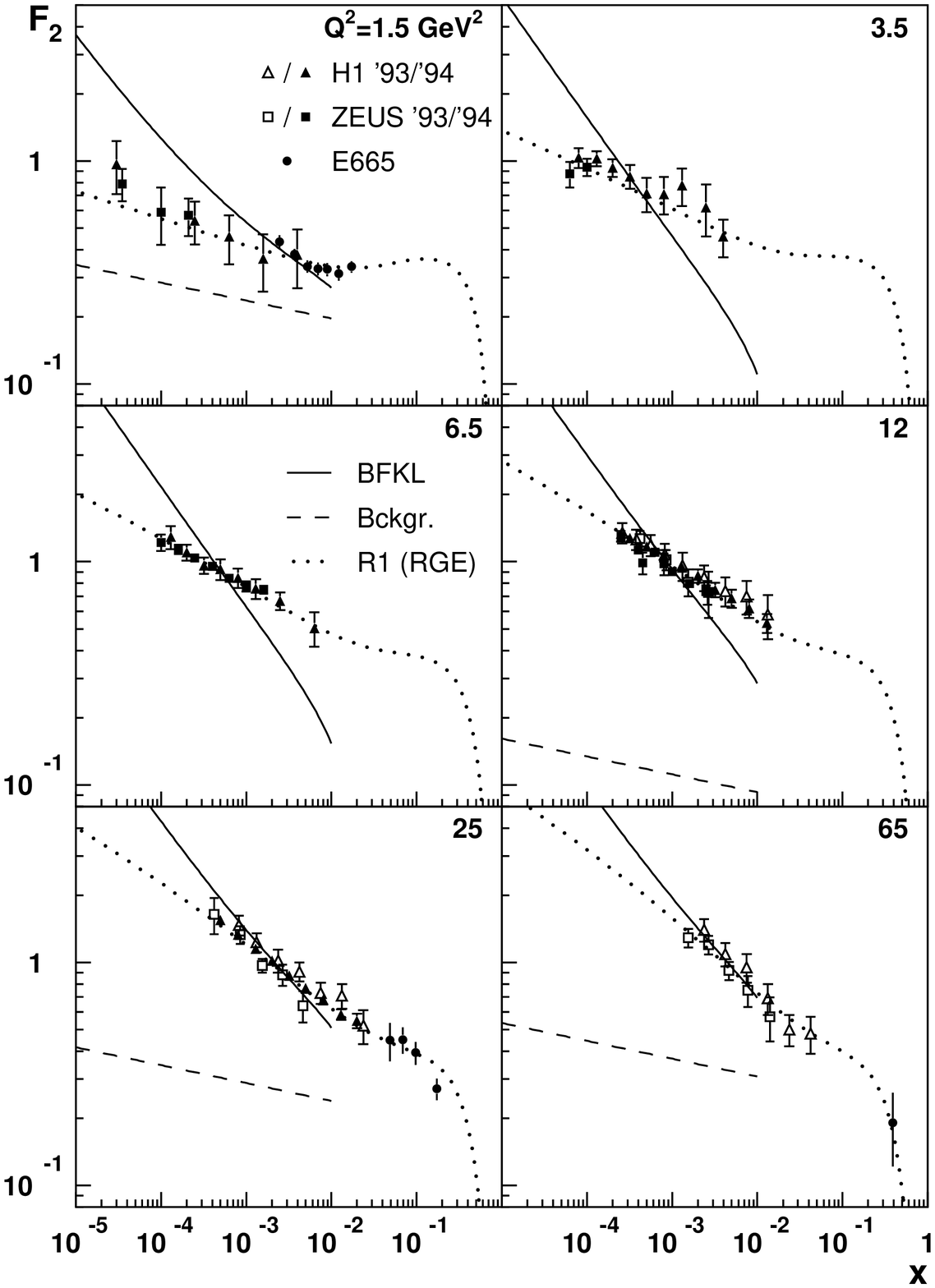,height=20cm,width=16cm}\\
\refstepcounter{figure}
\label{fig1}
\vspace{0.5cm}
{\large\bf Fig.\ \thefigure}
\end{center}
\end{figure}

\begin{figure}
\begin{center}
\epsfig{file=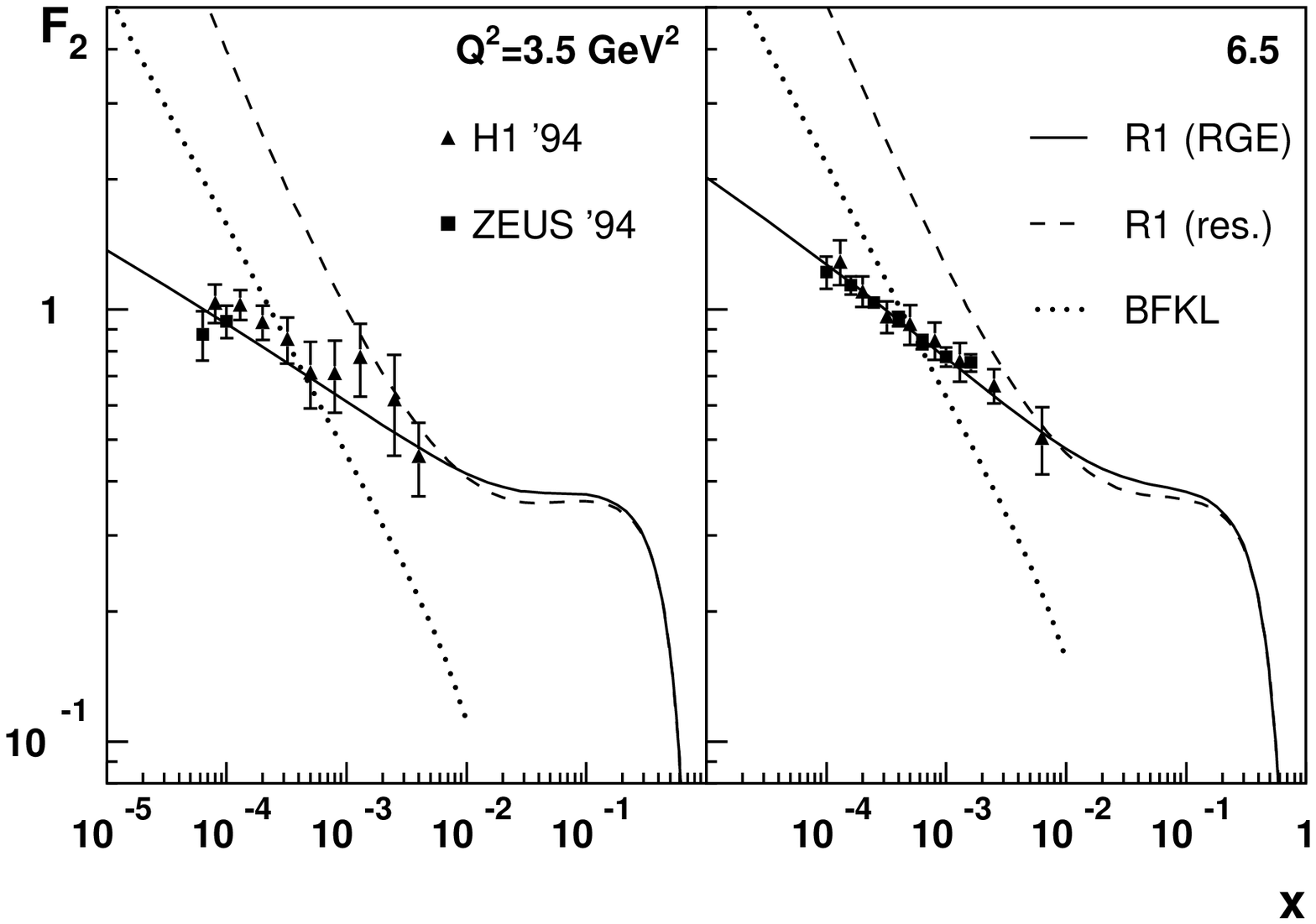,height=12cm,width=16cm}\\
\refstepcounter{figure}
\label{fig2}
\vspace{0.5cm}
{\large\bf Fig.\ \thefigure}
\end{center}
\end{figure}

\begin{figure}
\begin{center}
\epsfig{file=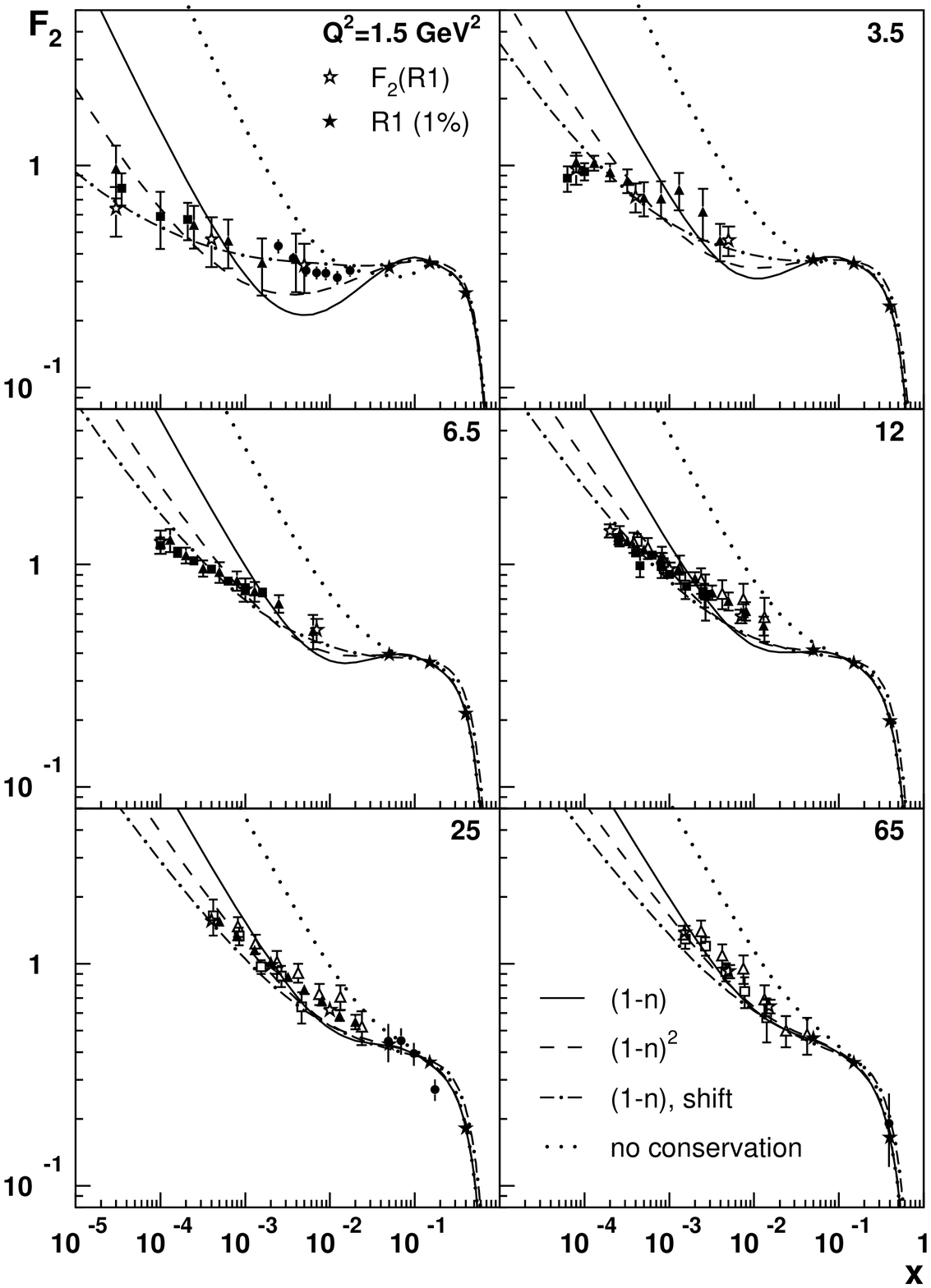,height=20cm,width=16cm}\\
\refstepcounter{figure}
\label{fig3}
\vspace{0.5cm}
{\large\bf Fig.\ \thefigure}
\end{center}
\end{figure}

\begin{description}
\item[Fig.\ \ref{fig1}]
BFKL predictions using the methods of \cite{oldpap} compared
with recent data \cite{HERA93,HERA94,E665}. The solid curve is
the LO BFKL prediction including the fitted background which is shown
separately as dashed curve. Only data with $x\le 10^{-2}$ is fitted.
The dotted curve shows the conventional RGE fit R1 of MRS \cite{MRSR}
for comparison.
\item[FIG.\ \ref{fig2}]
The original MRS-R1 input \cite{MRSR} evolved with the RGE (solid curve)
and the
RGE plus resummations (dashed curve). The resummations are added using
the conserving factor $(1-n)$. For comparison the LO BFKL prediction
is shown as dotted curve. Note that ``R1 (res.)'' is {\em not}
fitted to the shown data \cite{HERA94}.  
\item[Fig.\ \ref{fig3}]
Modified evolution predictions, based on the RGE supplemented by the small
$x$ resummations according to Eqs.\ (\ref{effcorr}) and (\ref{conadd}),
compared with experiment (data as in Fig.\ \ref{fig1}).
The stars show the generated data used for the fit
as explained in the text. The solid and dot-dashed curves are obtained
with $(1-n)$ as conserving factor, using a shifted $\alpha_s$ in the latter
case as discussed in the text, and the dashed curve is calculated with
$(1-n)^2$. The dotted
curve is obtained without any conserving factor, which corresponds
to the original results of \cite{catani}. 
\end{description}
%
%
%
\pagestyle{empty}
\vspace*{\fill}

\pagestyle{empty}
\vspace*{\fill}

\vspace*{\fill}
\end{document}